\documentclass[reprint,superscriptaddress,preprintnumbers,amsmath,amssymb,aps,prd]{revtex4-1}
\usepackage{mathtools}
\usepackage{color}
\usepackage[colorlinks=true,linkcolor=blue,citecolor=blue,urlcolor=blue]{hyperref}
\usepackage{graphicx}
\usepackage{dcolumn}
\usepackage{bm}
\usepackage{xcolor}
\usepackage{multirow}
\usepackage[normalem]{ulem}

\usepackage{tikz}
\newcommand{\subfig}[2]{%
\begin{tikzpicture}%
\node[rectangle] (image) at (0,0) {#2};
\node[anchor=south west] (label) at (image.south west) {(#1)};
\end{tikzpicture}%
}

\def\Fig#1{Fig.~\ref{#1}}
\def\Eq#1{Eq.~\eqref{#1}}

\def\Figs#1{Figs.~\ref{#1}}
\def\fig#1{\ref{#1}}

\begin{document}

\preprint{CERN-TH-2022-154}
\title{Ratios of jet and hadron spectra at LHC energies: measuring high-$p_T$ suppression without a $pp$ reference}

\author{Jasmine Brewer}
\author{Alexander Huss}
\author{Aleksas Mazeliauskas}
\author{Wilke van der Schee}
\affiliation{Theoretical Physics Department, CERN, CH-1211 Geneva 23, Switzerland}

\begin{abstract}

We analyze the reliability of several techniques for computing jet and hadron spectra at different collision energies.
This is relevant for discovering energy loss in the upcoming oxygen--oxygen (OO) run at the LHC, for which a reference $pp$ run is currently not planned.
For hadrons and jets we compute the ratio of spectra between different  $pp$ collision energies in perturbative QCD, which can be used to construct a reference spectrum.
Alternatively, it can be interpolated from measured spectra at nearby energies.
We estimate the precision of both strategies  for the spectra ratio relevant to the oxygen run,
and conclude that the central values agree to 4\% accuracy for hadrons and 2\% accuracy for jets.
Finally we propose taking the ratio of OO and $pp$ spectra at different collision energies, 
which cleanly separates the experimental measurement and theoretical computation.
\end{abstract}

\maketitle

\section{Introduction}
High-$p_T$ hadrons and jets in proton--proton ($pp$) collisions at the LHC are essential tools for the study of QCD~\cite{Butterworth:2012fj}.
In heavy-ion collisions, 
suppression of high momentum hadron and jet spectra is important evidence for the formation of strongly coupled Quark-Gluon Plasma~\cite{STAR:2005gfr,PHENIX:2004vcz,Aad:2010bu,Chatrchyan:2011sx,ALICE:2010yje}.
Intense searches for high-$p_T$ hadron or jet spectra suppression
in small collision systems, such as proton--lead, have so far been unsuccessful~\cite{CMS:2016xef,CMS:2016svx,ALICE:2018vuu,ATLAS:2016xpn, ATLAS:2014cpa,PHENIX:2019gix,PHENIX:2015fgy,ALICE:2017svf,ALICE:2014xsp} 
despite evidence for collective effects~\cite{CMS:2012qk,ATLAS:2012cix,ALICE:2012eyl,ALICE:2016fzo,LHCb:2015coe}.

The planned special oxygen--oxygen (OO) run at the LHC in Run 3~\cite{Citron:2018lsq} offers a unique opportunity to discover high-$p_T$ partonic rescatterings in small collision systems.
Several recent model studies indicate that energy loss effects can lead to the suppression of charged hadron spectra up to  $\sim 20\%$ for hadrons with momentum $\sim 20\,\text{GeV}$~\cite{Huss:2020dwe, Huss:2020whe,Zakharov:2021uza, Liu:2021izt}. The discovery potential therefore relies on the smallness of experimental uncertainties in the measurement and theoretical uncertainties in the reference baseline. In this paper we focus on the latter.

In principle, discovering final-state modification requires differentiating measured spectra for hard probes from the perturbative QCD (pQCD) expectation with nuclear parton distribution functions (nPDFs).
However both theoretical and experimental uncertainties on absolute hadron and jet spectra are substantial, so in practice it is crucial to cancel uncertainties by taking the ratio of spectra to a $pp$ reference. However, a measured $pp$ reference at the same energy might not be available in a short OO run~\cite{Brewer:2021kiv}.

In this work, we study the reliability and accuracy of several methods for measuring jet and hadron suppression in nuclear collisions without a measured $pp$ reference at the same energy.
We first consider constructing a $pp$ reference by rescaling a measured spectrum by a theoretically-computed ratio of spectra at different energies. We use the anticipated OO centre-of-mass energy ($6.37$~TeV) as a timely example~\cite{Citron:2018lsq,Brewer:2021kiv}.
Motivated by Ref.~\cite{ALICE-PUBLIC-2021-004} (see also Refs.~\cite{dEnterria:2004rur,Arleo:2010kw}), we compute these hadron and jet spectra ratios in pQCD and also extract them in a data-driven way from global fits of spectra at nearby collision energies.
Finally, we propose that jet or hadron energy loss can be studied by taking the ratio of OO and $pp$ spectra at \emph{different} collision energies and comparing to the corresponding ratio computed in pQCD. 
This method  disentangles the experimental measurement from theoretical input without significantly inflating the theoretical uncertainties on the baseline compared to the conventional nuclear modification factor.

\section{Perturbative QCD predictions}
\label{sec:pQCD}
\begin{figure*}
    \centering
\subfig{a}{\includegraphics[width=0.33\textwidth]{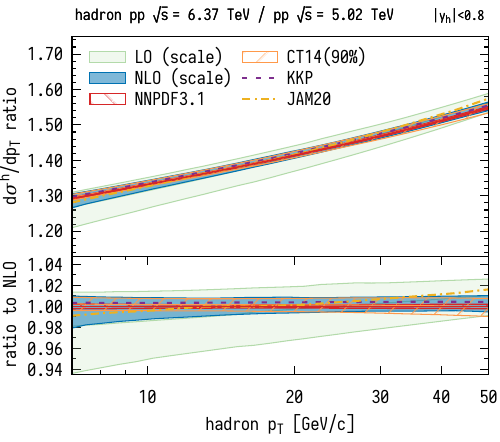}}%
\subfig{b}{\includegraphics[width=0.33\textwidth]{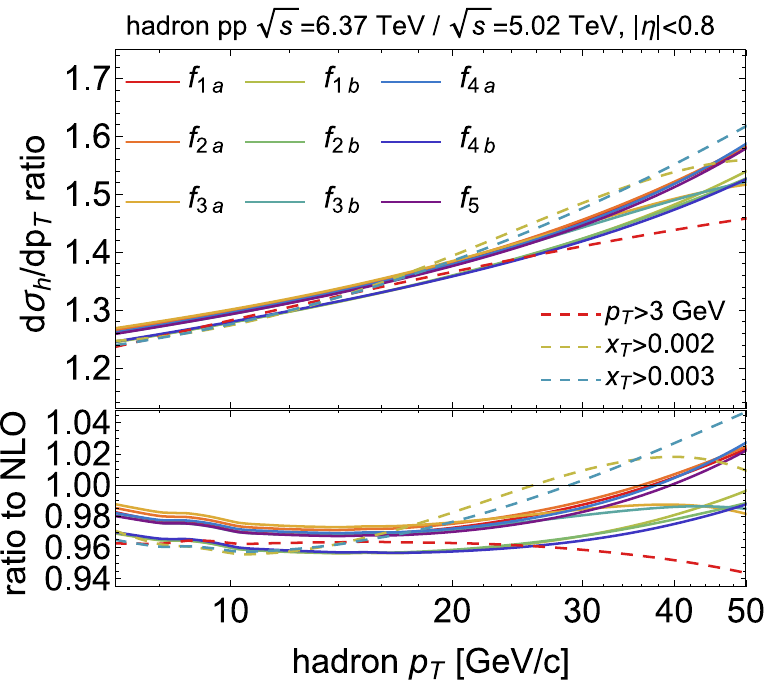}}%
\subfig{c}{\includegraphics[width=0.33\textwidth]{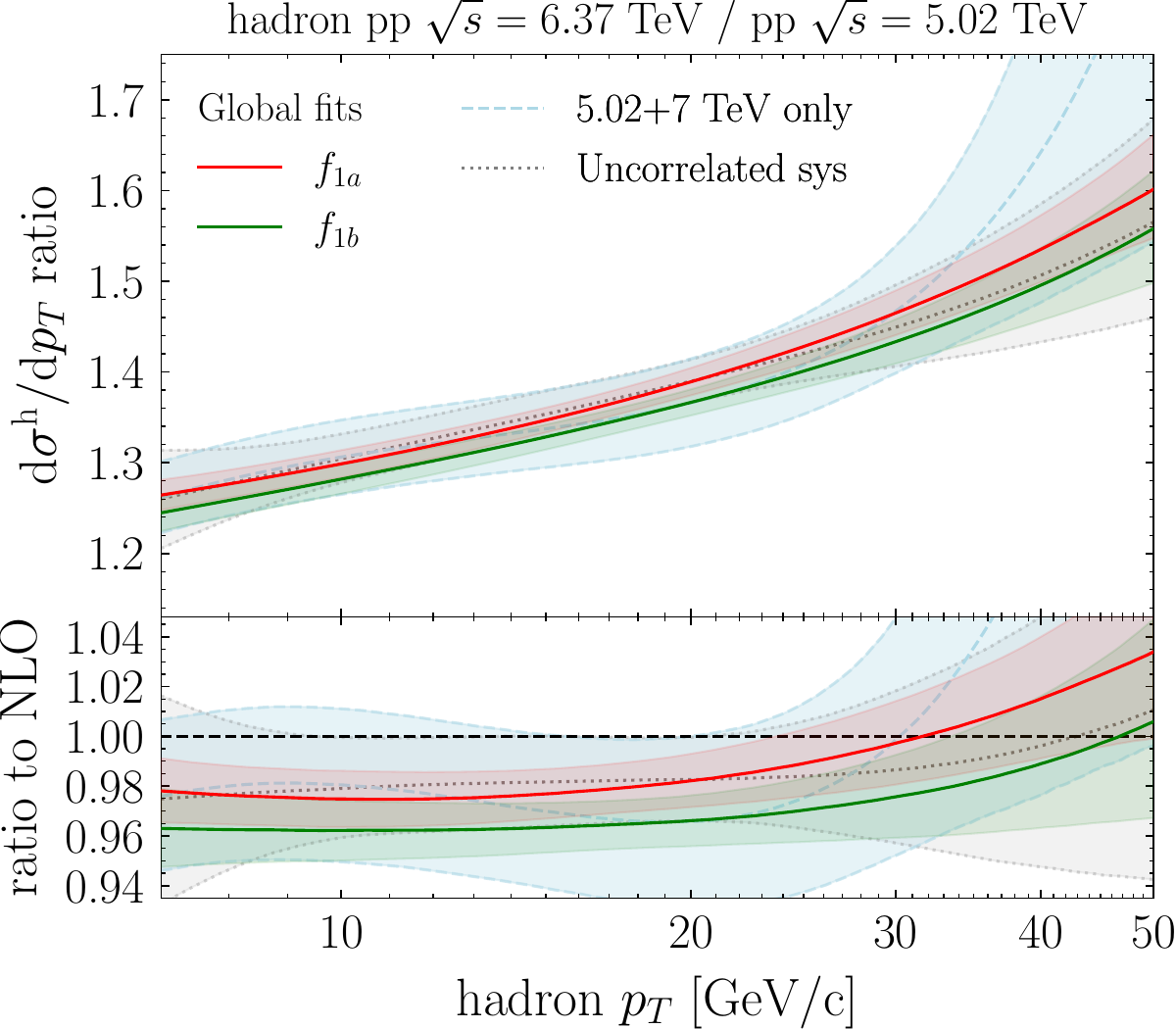}}
    \caption{Ratio of inclusive hadron spectra in $pp$ collisions at $\sqrt{s} = 6.37$ and 5.02 TeV. (a) Ratio from pQCD calculation. Bands show   LO (green) and NLO (blue) scale uncertainties and  NNPDF3.1 (red) and CT14 (orange) PDF uncertainties. Lines show additional hadron FFs (BKK is the default). (b) Ratio from the best fits using the forms in Table~\ref{tab:multiplefitsfull} to ALICE data \cite{Acharya:2018qsh,Abelev:2013ala} at 2.76, 5.02 and 7~TeV with $p_T>7\,\text{GeV}$ (solid) and other $p_T$ cuts for the form $f_{1a}$ (dashed). (c) Ratio from MCMC fits for $f_{1a}$ (red) and $f_{1b}$ (green) with 68\% confidence intervals (bands). Fits to only 5.02 and 7~TeV (blue) and using uncorrelated systematic uncertainties (grey) are also shown.
    }
    \label{fig1}
\end{figure*} 
\begin{figure*}
    \centering
\subfig{a}{\includegraphics[width=0.33\textwidth]{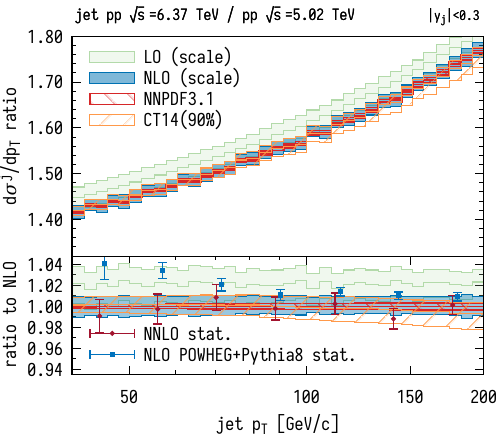}}%
\subfig{b}{\includegraphics[width=0.33\textwidth]{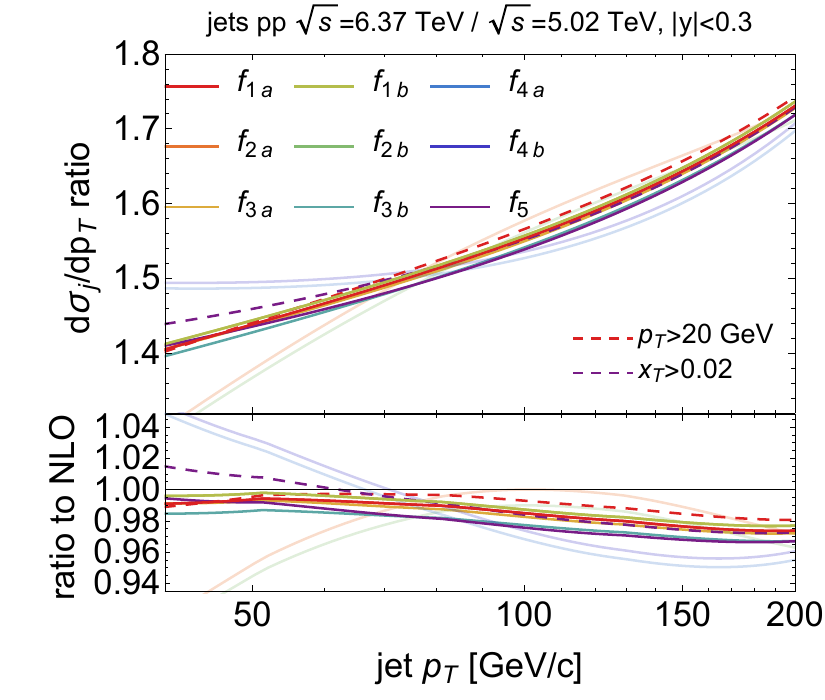}}%
\subfig{c}{\includegraphics[width=0.33\textwidth]{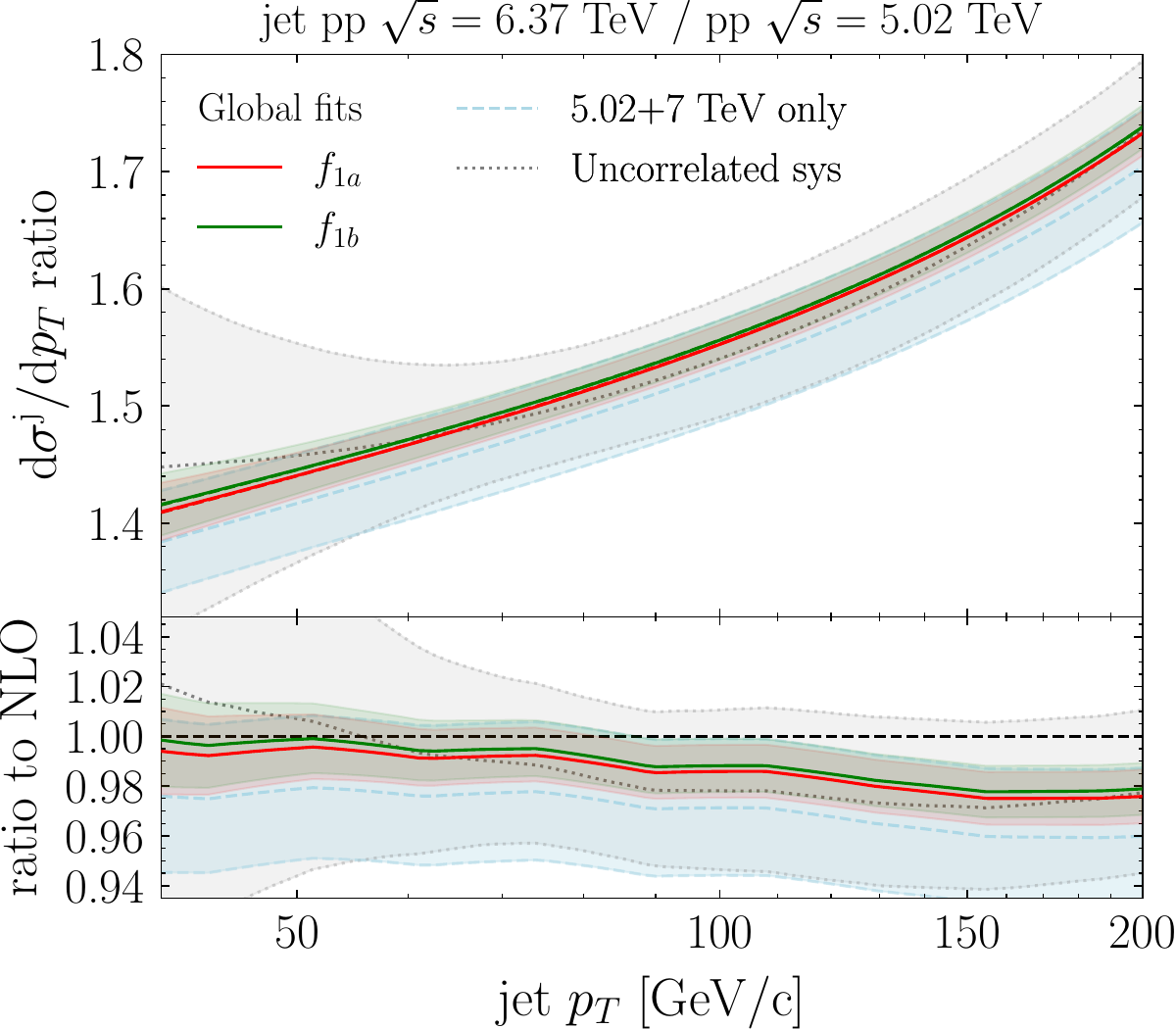}}
    \caption{As in Fig.~\ref{fig1} but for jet spectra. (a) The lower panel includes  NNLO partonic and showered NLO results. (b)  Solid lines show fits to ATLAS data~\cite{Aad:2011fc,Aaboud:2018twu,Aad:2013lpa} using a standard $p_T > 40\,$GeV cut and fit forms with $\chi _{\nu }^2>5$ are drawn with lower opacity.
    }
    \label{fig2}
\end{figure*} 
In \Fig{fig1}(a) and \Fig{fig2}(a) we present the ratios between  $\sqrt{s}=6.37$ and $5.02\,\text{TeV}$  inclusive hadron and jet spectra in $pp$ collisions computed in pQCD.
In collinear QCD factorization, the non-perturbative long distance part of the cross-section---the parton distribution functions (PDFs) and hadron fragmentaton functions (FFs)---are factorized at a characteristic scale $\mu_F$ from the the hard partonic scattering cross-section.
Perturbative cross-sections depend on the unphysical factorization  and renormalization scales $\mu_F$ and $\mu_R$, which are commonly chosen to represent a typical hard scale of the process. For hadrons 
we choose the central scales to be the hadron momentum $\mu_0=p_T^h$, while for jets we used the scalar sum of the transverse momenta of all partons $\mu_0 = \sum_{i} p_T^i$~\cite{Currie:2018xkj}. We estimate uncertainties due to missing higher-order corrections by varying $\mu_F$ and $\mu_R$ around $\mu_0$ by factors of $2$ and $1/2$ while imposing the restriction $1/2\leq \mu_F/\mu_R\leq 2$. We vary $\mu_F$ for PDFs and FFs independently. 

The $p_T$ differential hadron cross-section was computed at leading order (LO) and next-to-leading order (NLO) using a modified version of the INCNLO code~\cite{Aversa:1988vb}\footnote{\url{http://lapth.cnrs.fr/PHOX_FAMILY/readme_inc.html}} with LHAPDF grid support~\cite{Buckley:2014ana}. For the charged hadrons we used BKK FFs~\cite{Binnewies:1994ju} (the sum of pion and kaon FFs) as the default choice, but we also computed central values using KKP~\cite{Kniehl:2000fe} and JAM20~\cite{Moffat:2021dji} FFs.
Jet spectra at LO, NLO, and next-to-next-to-leading order (NNLO) were computed using the \textsc{NNLOjet} framework with the anti-$k_T$ algorithm with a jet radius $R=0.4$~\cite{Currie:2016bfm,Gehrmann:2018szu} and using \texttt{APPLfast} interpolation tables~\cite{Britzger:2019kkb}.
We evaluated uncertainties due to proton PDFs and show for NNPDF3.1~\cite{NNPDF:2017mvq} the $\pm1\sigma$ uncertainty band, while for CT14~\cite{Dulat:2015mca} we display the 90\% confidence level band.

In the lower panel of \Fig{fig1}(a) we see that owing to the cancellation in the correlated scale variation for the ratio between the two energies, the NLO scale envelope  for hadrons is less than 2\% in the momentum range $p_T^h \in [7,50]\,\text{GeV}$ and contained within the LO scale band. For jets in \Fig{fig2}(a) the NLO scale envelope is below 1\% for $p_T^j \in  [40,200]\,\text{GeV}$ and the central NNLO prediction agrees with it within statistical uncertainties. We also observe excellent cancellation of PDF uncertainties with negligible uncertainties using the NNPDF3.1 set, which are fully consistent with the CT14 error band for both jet and hadron spectra. 
For hadrons FFs are consistent within 2\% accuracy.
The showered jet spectra are consistent with the partonic ratio within 2\% for $p_T>60\,\text{GeV}$.
We checked that the deviation at low $p_T$ is due to the $\sqrt{s}$ dependence of hadronization effects (see below)~\cite{Aad:2013lpa}.

\begin{figure*}
    \centering
    \subfig{a}{\includegraphics[width=0.4\linewidth]{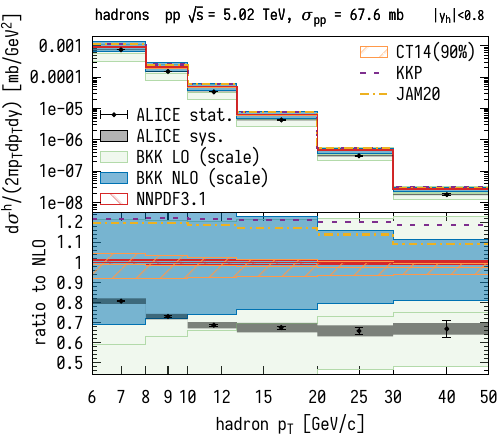}}
    \subfig{b}{\includegraphics[width=0.4\linewidth]{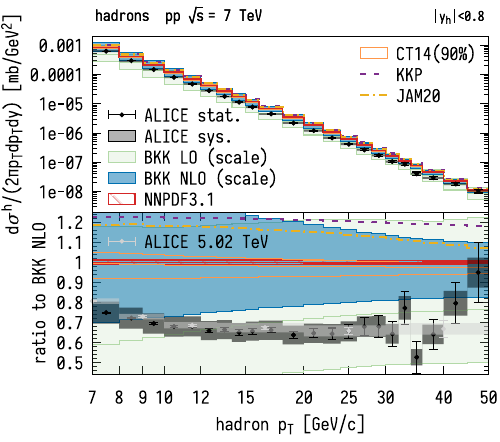}}
    \caption{Differential hadron spectra at (a) 5.02~TeV and (b) 7~TeV by ALICE~\cite{Acharya:2018qsh,Abelev:2013ala} compared to pQCD predictions. Absolute spectra (upper panels) and ratio to NLO (lower panels) are shown. Scale uncertainties are shown by solid bands, PDF uncertainty by patterned bands, and additional FFs by lines (BKK is the default). The ratio of 5.02~TeV data to NLO is overlaid in light grey in the lower panel of (b).
	}
    \label{fig:datatopQCDhadrons}
\end{figure*}

\begin{figure*}
    \centering
    \subfig{a}{\includegraphics[width=0.4\linewidth]{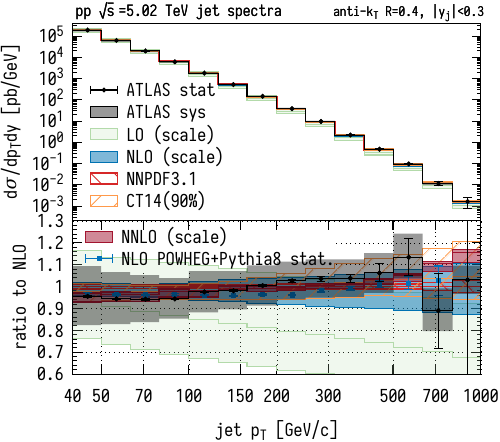}}
    \subfig{b}{\includegraphics[width=0.4\linewidth]{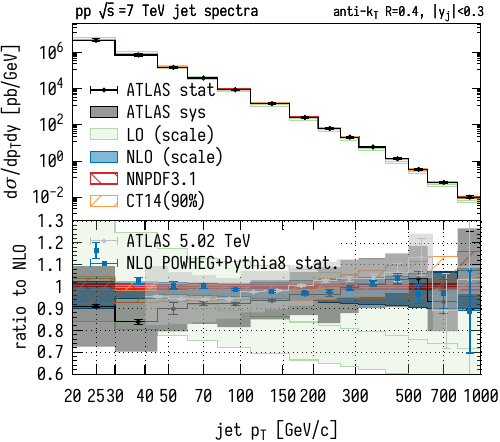}}
    \caption{Differential jet spectra at (a) 5.02~TeV and (b) 7~TeV by ATLAS~\cite{Aad:2011fc,Aaboud:2018twu} compared to pQCD predictions. Absolute spectra (upper panels) and ratio to NLO (lower panels) are shown. Scale uncertainties are shown by solid bands and PDF uncertainty by patterned bands. NNLO scale band and showered NLO results are shown in the lower panel of (a). The ratio of 5.02~TeV data to NLO is overlaid in light grey in the lower panel of (b). }
    \label{fig:datatopQCDjets}
\end{figure*}

We have tested the pQCD computation by direct comparison with existing $pp$ measurements at $\sqrt{s}=5.02$ and 7 TeV for hadrons by ALICE~\cite{Acharya:2018qsh,Abelev:2013ala} (\Fig{fig:datatopQCDhadrons}) and jets by ATLAS~\cite{Aad:2011fc,Aaboud:2018twu} (\Fig{fig:datatopQCDjets}). The upper panels show the absolute differential spectra, while the lower panels display the ratio to the NLO pQCD prediction (central scale).


The hadron spectra are systematically overpredicted and experimental points are mostly outside the NLO scale band. The large discrepancy in reproducing hadron spectra in hadron collisions is a well known problem of fragmentation functions that are fitted solely to $e^+e^-$ data~\cite{dEnterria:2013sgr}.
Despite the discrepancies in the absolute cross-section, we note that the data to theory ratio with BKK FFs is remarkably stable for $\sqrt{s}=5.02$ and 7~TeV. To see this we overlay the $\sqrt{s}=5.02$~TeV ratio to NLO in light grey in the lower panel of  \Fig{fig:datatopQCDhadrons}(b), which is consistent with the ratio for $7$~TeV within experimental uncertainties.
Therefore we note that the systematic deviations largely cancel between two energies and pQCD predicts a $5.02/7$~TeV ratio that is consistent with data.

The jet spectrum at 5.02~TeV in \Fig{fig:datatopQCDjets}(a) is reproduced within NLO scale uncertainties with a slightly rising data to theory ratio that is captured by NNLO corrections. The non-perturbative and showering effects of jet spectra were tested at NLO using POWHEG+Pythia8~\cite{Alioli:2010xd,Sjostrand:2014zea}%
\footnote{POWHEG Events were generated using the jet $p_T$ of the underlying Born kinematics as the central scale (default) and the settings \texttt{bornzerodamp=1} and \texttt{hdamp=100}.} and were found to be consistent with fixed-order NLO predictions within scale uncertainties. \Fig{fig:datatopQCDjets}(b) shows a similar plot for 7~TeV data (with 5.02~TeV overlaid in light grey). Again we see that the data to NLO ratio changes little between the two energies and the showered spectra at NLO are consistent with data down to 40~GeV. 
This was also seen in the ATLAS analysis of jets at $2.76$ and $7$~TeV~\cite{Aad:2013lpa}.

\section{Interpolation of experimental data}
\label{sec:interpolation}
Here we obtain the ratios of inclusive hadron and jet spectra in $pp$ collisions at $\sqrt{s}=6.37$ and $5.02\,\text{TeV}$ from interpolation of measured data at nearby collision energies. Interpolation was used to construct references for 5.02~TeV proton--lead run~\cite{ALICE:2012mj, ATLAS:2014cpa, ATLAS:2016xpn, CMS:2015ved, CMS:2016svx}
(see also \cite{dEnterria:2004rur,Arleo:2010kw} and \cite{D0:2020dsd}).
A first look at the interpolated hadron spectra ratio for oxygen was reported in a recent ALICE public note~\cite{ALICE-PUBLIC-2021-004}. Here we present a detailed study of the uncertainties arising from interpolation, both by studying the dependence on the functional form of the fits (Fig.~\ref{fig1}(b)  and  Fig.~\ref{fig2}(b))  and  using  Markov  Chain Monte Carlo (MCMC) fits to propagate data uncertainties into uncertainties on the spectra ratio (Fig.~\ref{fig1}(c) and Fig.~\ref{fig2}(c)).
  
To demonstrate the feasibility of these techniques we will use existing experimental data from LHC Run 1 and 2 at $\sqrt{s}=2.76$, 5.02, and 7~TeV. This is only illustrative, since $pp$ references taken during Run 3 will be at different energies (e.g., $5.02$, $8.8$, and $13.6$~TeV)~\cite{Citron:2018lsq}.
For charged hadrons we use differential cross-sections with $|\eta|<0.8$ measured by ALICE~\cite{Acharya:2018qsh,Abelev:2013ala}. At $\sqrt{s}=2.76$ and $5.02$~TeV we convert measured spectra to cross-sections using inelastic hadronic cross-sections of $61.8$ and $67.6\,$mb, respectively~\cite{Loizides:2017ack}.
We use ATLAS measurements of inclusive jet spectra with $|y|<0.3$ and anti-$k_T$ radius $R=0.4$~\cite{Aad:2011fc,Aaboud:2018twu,Aad:2013lpa}.
Unless otherwise stated, we fit only data satisfying $p_T>7$~GeV for hadrons and $p_T>40$~GeV for jets.
In the appendix~\ref{sec:appA}, we also show results for CMS measurements of hadron spectra at the same energies \cite{CMS:2011mry, CMS:2012aa,CMS:2016xef}.

We obtain the best fits by maximizing the log likelihood function (or minimizing $\chi^2 \equiv -2 \log \mathcal L$), with
\begin{align}
\label{eq:lnlike}
    \log \mathcal{L} = &-\frac{1}{2}\sum_{\sqrt{s},i,j} \Delta y^{\sqrt{s}}_i (C^{\sqrt{s}})^{-1}_{ij} \Delta y^{\sqrt{s}}_j,\\
    \Delta y_i^{\sqrt{s}} = &y^{\sqrt{s}}_i - \int_{\text{bin i}} \text{d} p_T \, f(\sqrt{s},p_T).
\end{align}
Here 
$y_i^{\sqrt{s}}$
is the $i$-th datapoint at $\sqrt{s}$,
$f(\sqrt{s},p_T)$ is a fit to the differential spectrum,
and $C^{\sqrt{s}}$ is the covariance matrix for that collision energy.
In the absence of the available covariance matrix of experiment uncertainties, we model $C^{\sqrt{s}}$ with two limiting cases. We treat $C^{\sqrt{s}}$ as a sum of uncorrelated uncertainties $\sigma_\text{uncorr}^{\sqrt{s}}$ and a luminosity uncertainty $\sigma_\text{lum}^{\sqrt{s}}$ that is fully correlated at each energy, 
\begin{equation}
\label{eq:sigma}
(C^{\sqrt{s}})_{ij} = (\sigma^{\sqrt{s}}_{\text{uncorr},i})^2 \delta_{ij} +(\sigma^{\sqrt{s}}_\text{lum})^2.
\end{equation}
In the first case, we consider that all systematic uncertainties apart from the luminosity uncertainty cancel in the ratio between energies, so the uncorrelated uncertainty is only statistical.
For comparison we also consider the opposite limit that all non-luminosity systematic uncertainties are fully uncorrelated and add them in quadrature to the statistical uncertainties.
We expect that these two scenarios bracket the actual covariance matrix of experimental uncertainties, which, we hope, will be provided with future measurements.

  \begin{table*}
 \caption{Fitting forms together with the resulting reduced $\chi^2$ for fits of hadron ($p_T > 7\,$GeV) and jet ($p_T > 40\,$GeV) spectra for $pp$ collisions at $\sqrt{s}=2.76,\,5.02\,$and$\,7\,$TeV. Here $\hat{x}_T = x_T/x_0$ with $x_0=0.003$ for hadrons and $x_0=0.02$ for jets. $a$, $b$, $c$ and $d$ are linear functions of either $\sqrt{\tilde{s}}\equiv \sqrt{s}/2.76\,\text{TeV}$ or $\log \sqrt{\tilde{s}}$, e.g. $a(\tilde{s}) = a_0 + a_1 \sqrt{\tilde{s}}$ or $a(\tilde{s}) = a_0 + a_1 \log \sqrt{\tilde{s}}$.  
 All other variables are fit parameters.
 \label{tab:multiplefitsfull}
 }
 \begin{tabular}{lllcccc}
 \hline
 \hline
  function & fit form & $\sqrt{s}$ dep. & p & $\chi _{\nu }^2\text{-hadrons}$ & $\chi_{\nu }^2\text{-jets}$ \\
  \hline
  $f_{1a}$ & \multirow{2}{*}{$A\left(\sqrt{\tilde{s}}\right)^{\beta } \hat{x}_T^{a(\tilde{s})+b(\tilde{s}) \hat{x}_T+c(\tilde{s}) \log \hat{x}_T} $} &linear & 8 & 0.75 & 1.11 \\
  $f_{1b} $& & log linear &8 & 0.75 & 1.15 \\
\hline
  $f_{2a} $& \multirow{2}{*}{$A\left(\sqrt{\tilde{s}}\right)^{\beta } x_T^{a(\tilde{s})+b(\tilde{s}) x_T+c(\tilde{s}) \log x_T} $}& linear & 8 & 0.75 & 5.02 \\
 $ f_{2b}$ & &log linear& 8 & 0.75 & 5.10 \\
\hline
  $f_{3a}$ & \multirow{2}{*}{ $A\left(\sqrt{\tilde{s}}\right)^{\beta } \hat{x}_T^{a(\tilde{s})+b(\tilde{s}) \hat{x}_T+c(\tilde{s}) \log \hat{x}_T +d(\tilde{s})\hat{x}_T/ \log \hat{x}_T } $} & linear & 10 & 0.80 & 1.19 \\
  $f_{3b}$ & &log linear& 10 & 0.80 & 1.18 \\
\hline
 $ f_{4a}$ & \multirow{2}{*}{$A\left(\sqrt{\tilde{s}}\right)^{\beta } \hat{x}_T^{a(\tilde{s})+b(\tilde{s}) \hat{x}_T} $} &linear & 6 & 0.70 & 5.05 \\
 $f_{4b}$ &  &log linear& 6 & 0.70 & 5.10 \\
 \hline
 $ f_5$ & $A\left(\sqrt{\tilde{s}}\right)^{\beta } \hat{x}_T^{a_0+a_1 \sqrt{\tilde{s}}+a_2 \hat{x}_T +a_3 \log(\sqrt{\tilde{s}}) +a_4 \log \left(\hat{x}_T\right)}$ &--& 7 & 0.73 & 1.10 \\
  \hline
 \hline
 \end{tabular}
 \end{table*}

The high-$p_T$ differential cross-section of hadrons or jets can be approximated by a power of $\sqrt{s}$ times a function of the dimensionless variable $x_T = 2 p_T/\sqrt{s}$
\cite{Brodsky:1973kr,Sivers:1975dg,ATLAS:2013pbc}.
We therefore perform global fits of either hadron or jet spectra 
using one of nine fitting formulas of the form
\begin{equation}
\label{eq:dsigma}
    \frac{d\sigma}{dp_T} =f(\sqrt{s},p_T) = A \sqrt{\tilde{s}}^\beta \hat{x}_T^{n(\hat x_T, \sqrt{\tilde{s}})}.
\end{equation}
Here $A$ is a dimensionful normalization constant, $\sqrt{\tilde{s}}\equiv \sqrt{s}/2.76\,\text{TeV}$ 
and $\hat{x}_T\equiv x_T/x_0$, with $x_0=0.003$ ($x_0=0.02$) for hadrons (jets). 
We summarize different parametrizations in
Table~\ref{tab:multiplefitsfull}. The first eight forms come in pairs, with the only difference being the $\sqrt{\tilde{s}}$ dependence of the coefficients in the exponent $n(\hat x_T, \sqrt{\tilde{s}})$, which we assume to be either linear in $\sqrt{\tilde{s}}$ or in $\log\sqrt{\tilde{s}}$.
The form $f_2$ differs from $f_1$ in using a power of $x_T$ instead of $\hat{x}_T$. Restricted to 2.76 and 7~TeV data, the $f_{2b}$ fit is equivalent to the method  used in \cite{Aad:2013lpa,ATLAS:2014cpa}. 
We calculate the reduced $\chi^2_\nu = \chi^2/(N-p)$ for each fit form in Table~\ref{tab:multiplefitsfull}, with $N$ and $p$ the number of data points and parameters. For default $p_T$ cuts, $N=32$ for hadrons and $37$ for jets.
Our preferred fit forms $f_{1a,1b}$  with eight independent parameters have  $\chi^2_\nu\approx0.75$ and $1.1$ for hadrons and jets respectively, indicating a good quality of fits.
For jet spectra using $x_T$ instead of $\hat{x}_T$ as in $f_{2a,2b}$,  or reducing the number of fit parameters to 6 as in $f_{4a,4b}$ considerably worsens the fit quality. Increasing fit parameters to 10 in $f_{3a,3b}$ or considering mixed $\sqrt{\tilde{s}}$ dependence  in $f_5$ left $\chi^2_\nu$ for jets essentially unchanged. For hadrons all nine fit forms give comparable fit quality.

We use the interpolation functions to compute the jet and hadron spectra ratios at $\sqrt{s}=6.37$ and $5.02\,\text{TeV}$ shown in \Figs{fig1}(b) and \fig{fig2}(b). For hadrons the fit forms with linear $\sqrt{\tilde s}$ dependence result in about 2\% larger ratio compared to $\log\sqrt{\tilde s}$. For jets there is no strong dependence on the $\sqrt{\tilde{s}}$ parametrization. For jets fit forms $f_{2a,2b,4a,4b}$ did not lead to a satisfactory $\chi^2_\nu$ and are hence drawn at lower opacity in the figure.

We also investigate our fit sensitivity to data selection using the $f_{1a}$ fit form in \Figs{fig1}(b) and \fig{fig2}(b).
 The lower $p_T$ (or $x_T$) cut on data is necessary for hadrons, which are measured to much lower $p_T$ than $x_T$ scaling applies.
For hadrons we consider $x_T>0.003$ ($\chi_{\nu }^2=1.14$), $p_T > 3\,$GeV ($\chi_{\nu }^2=1.72$), and $x_T>0.002$ ($\chi_{\nu }^2=1.60$). 
For the latter two, the fit is not consistent with the data if all systematics cancel ($\chi_{\nu }^2=16.2$ and $\chi_{\nu }^2=8.7$) so we include 1\% uncorrelated systematic uncertainties in these fits.
Lowering the $p_T$ cut to 3~GeV changes the hadron ratio at higher momentum, but with worse $\chi^2_\nu$.
For jets we show results for $p_T>20\,$GeV (all available data) and $x_T>0.02$ with $\chi _{\nu }^2=3.36$ and $1.16$.
Jet fits are relatively insensitive to the data selection.

Next we study the propagation of experimental uncertainties to the interpolated ratio. We use $f_{1a,1b}$ (again with $p_T>7$~GeV for hadrons and $p_T>40$~GeV for jets) and perform MCMC global fits using the \textsf{emcee} implementation~\cite{ForemanMackey:2012ig}. 
We use the likelihood function \Eq{eq:lnlike} to sample the distribution of parameters consistent with the data at the anchor energies given their uncertainty covariance matrix (from \Eq{eq:sigma}).
In particular, we initialize a chain of $200$ walkers with a Gaussian distribution in the parameter space around the maximum likelihood point.  The walkers are evolved according to the likelihood function \Eq{eq:lnlike} for $10^4$ steps with no constraints in the parameter space, i.e., with uniform priors. We use only the last $2000$ steps for the computation of the confidence intervals. Crucially, this method produces a distribution of fits obeying the likelihood function with experimental uncertainties, rather than a single fit that maximizes the likelihood. We then compute the 68\% confidence intervals on the interpolated ratio around the median value from the distribution of fit values in each momentum bin.

\begin{figure}
    \centering
    \includegraphics[width=\linewidth]{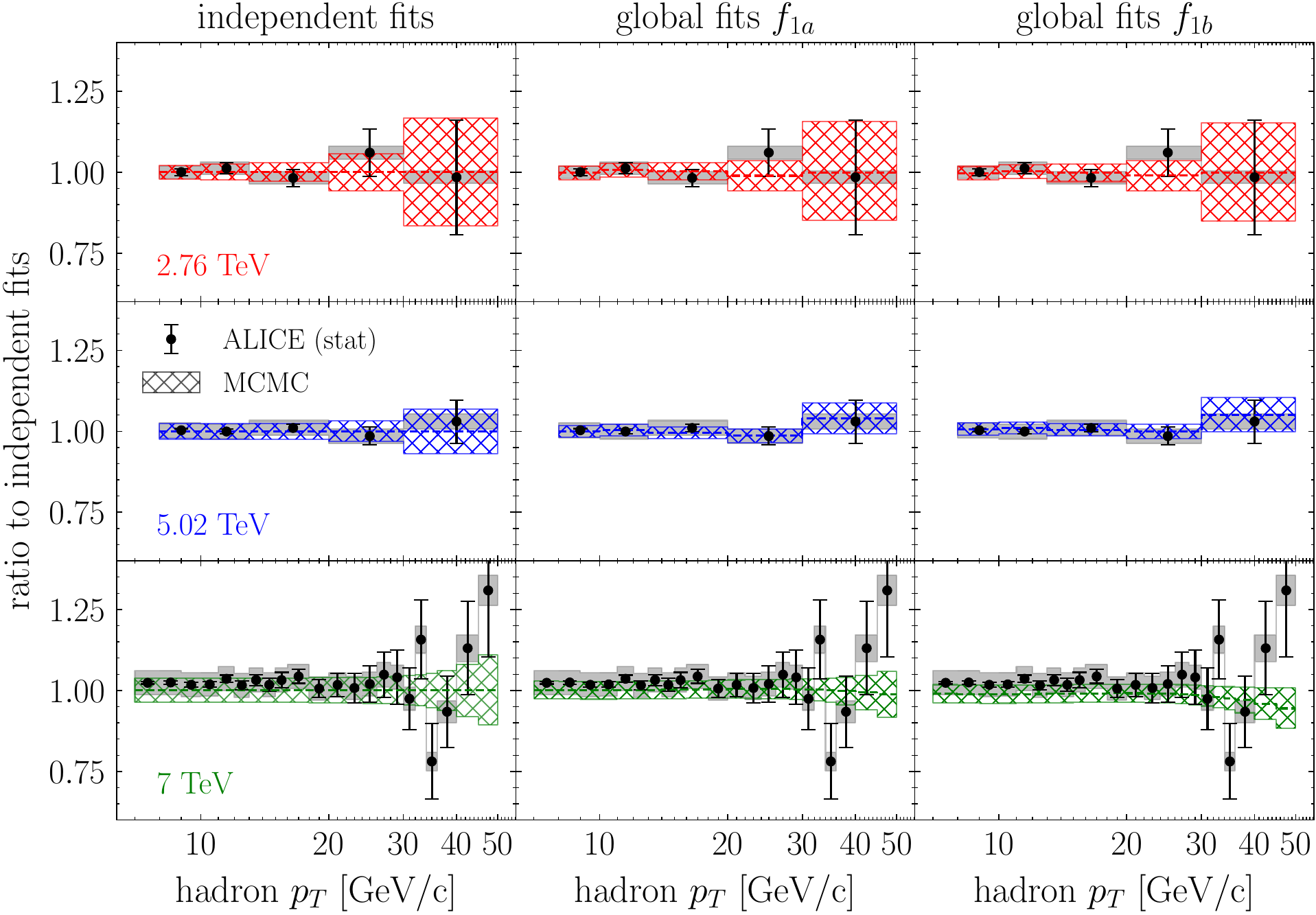}
    \caption{\label{fig:hadron_fits_lumionly}Fits to ALICE hadron spectra at $\sqrt{s}=2.76$, 5.02, and 7~TeV~\cite{Aad:2011fc,Aaboud:2018twu}, anchoring the interpolation for the results in \Fig{fig1}c. The left panel shows independent fits to the spectrum at each energy normalized to the median value of the independent fit at that energy. The center and right panels show results for global fits $f_{1a}$ and $f_{1b}$ normalized to the median value of the independent fit at that energy.  Uncorrelated (statistical) uncertainties are shown in bars and luminosity uncertainties are grey boxes. Hatched bands indicate the 68\% confidence interval of the MCMC fits.}
\end{figure}

\begin{figure}
    \centering
    \includegraphics[width=\linewidth]{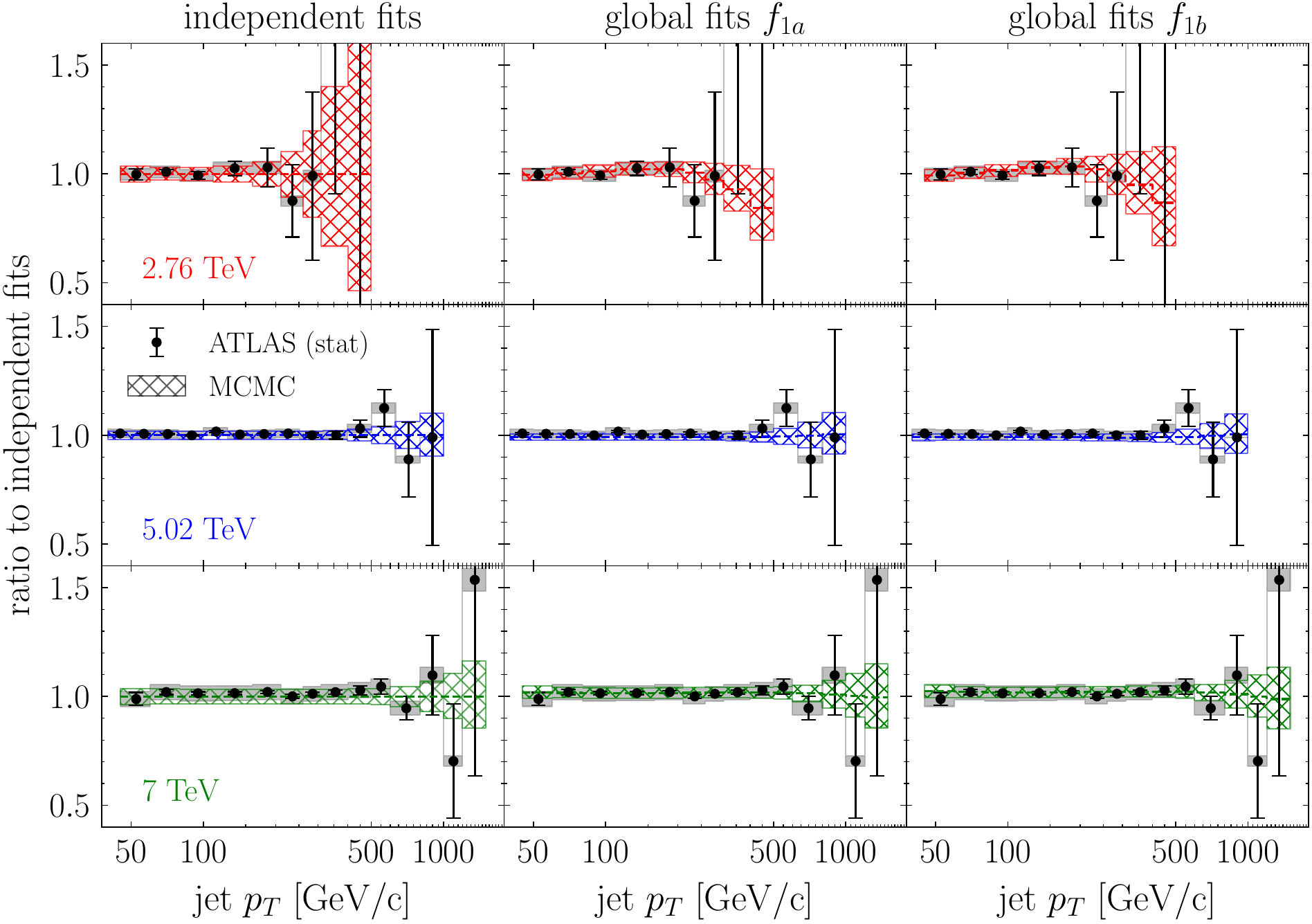}
    \caption{\label{fig:jet_fits_lumionly}Fits to ATLAS jet spectra at $\sqrt{s}=2.76$, 5.02, and 7~TeV~\cite{Aad:2011fc,Aaboud:2018twu,Aad:2013lpa}, anchoring the interpolation for the results in \Fig{fig2}c. The left panel shows independent fits to the spectrum at each energy normalized to the median value of the independent fit at that energy. The center and right panels show results for global fits $f_{1a}$ and $f_{1b}$ normalized to the median value of the independent fit at that energy.  Uncorrelated (statistical) uncertainties are shown in bars and luminosity uncertainties are grey boxes. Hatched bands indicate the 68\% confidence interval of the MCMC fits.}
\end{figure}

\Figs{fig1}(c) and \fig{fig2}(c) show the median and the $68\%$ confidence bands for the $6.37/5.02$~TeV spectra ratios for hadrons and jets. For hadrons, statistical and luminosity uncertainties on the spectra at the anchor energies give rise to few-percent uncertainties for $p_T \lesssim 30$~GeV that grow to $\sim 5\%$ at 50~GeV. $f_{1a}$ and $f_{1b}$ are consistent within uncertainty bands but shifted from each other at the 2\% level. Both fit forms deviate from the NLO pQCD prediction by a few percent for $p_T \lesssim 20$~GeV. For jets the results for $f_{1a}$ and $f_{1b}$ are very similar with $\sim 2\%$ uncertainties for all $p_T$ shown. The interpolated ratio has up to 2\% deviations from NLO pQCD for $p_T \gtrsim 120$~GeV. For both hadrons and jets, fits using only $5.02$ and $7$~TeV data have much larger uncertainties than fits of three energies. If non-luminosity systematic uncertainties are uncorrelated and not cancelled in the ratio, the median $6.37/5.02$~TeV ratio is similar but with a significant increase in uncertainties.
In the appendix~\ref{sec:appA} we show the same analysis for CMS hadron data, which has milder $p_T$ dependence of the ratio but is consistent within uncertainties.

\begin{figure}
    \centering
    \includegraphics[width=\linewidth]{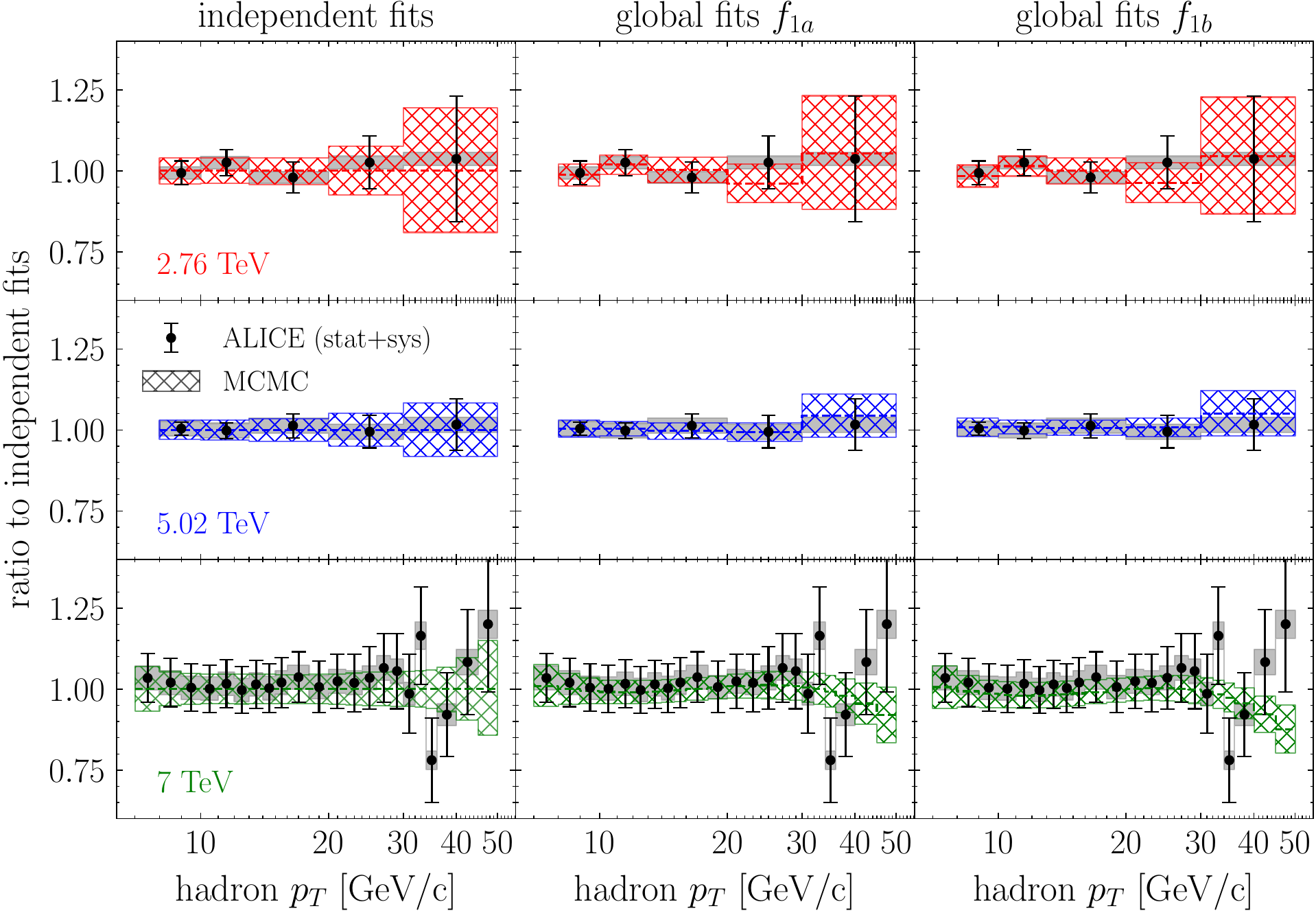}
    \caption{\label{fig:hadron_fits_uncorr} Same as \Fig{fig:hadron_fits_lumionly}, but assuming uncorrelated systematic uncertainties. Uncorrelated (statistical + systematic) uncertainties are shown in bars and luminosity uncertainties are grey boxes.}
\end{figure}

\begin{figure}
    \centering
    \includegraphics[width=\linewidth]{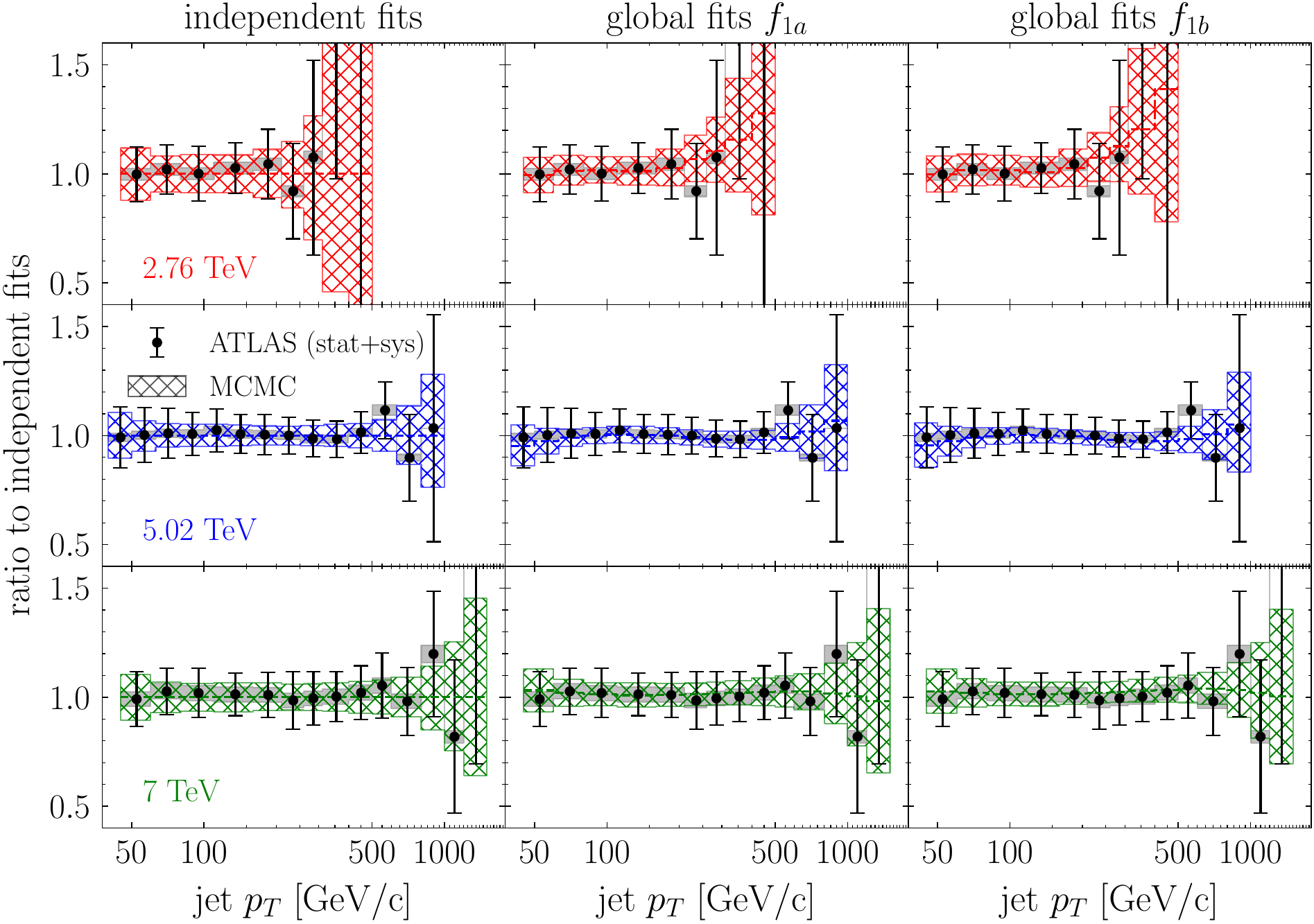}
    \caption{\label{fig:jet_fits_uncorr} Same as \Fig{fig:jet_fits_lumionly}, but assuming uncorrelated systematic uncertainties. Uncorrelated (statistical + systematic) uncertainties are shown in bars and luminosity uncertainties are grey boxes.}
\end{figure}

We test the MCMC global fits by showing the extracted  confidence intervals compared to measured spectra at the anchor energies.
We benchmark the global fits against fits to the individual spectra using the forms $f_1$ at a single center-of-mass energy (taking $a_1=0$ in Table~\ref{tab:multiplefitsfull}), in which case $f_{1a}=f_{1b}$.  We show the independent fits (left panel) and global fits $f_{1a}$ and $f_{1b}$ (center and right panels, respectively) for hadrons (\Fig{fig:hadron_fits_lumionly}) and for jets (\Fig{fig:jet_fits_lumionly}), assuming that systematic uncertainties besides the luminosity uncertainty cancel fully between energies. For visual clarity we show both data and fits as a ratio to the central value of the independent fit. \Figs{fig:hadron_fits_uncorr} and \ref{fig:jet_fits_uncorr} show the same information except assuming that non-luminosity systematic uncertainties are fully uncorrelated.
We note that in \Figs{fig:hadron_fits_uncorr} and \ref{fig:jet_fits_uncorr} the MCMC bands for the independent fits do not encompass the uncertainties (particularly obvious at 7~TeV in \Fig{fig:hadron_fits_uncorr} and 5.02~TeV in \Fig{fig:jet_fits_uncorr}), suggesting that the all systematic uncertainties cannot be considered fully uncorrelated between neighboring $p_T$ bins.

\section{Mixed-energy nuclear modification factor}
\label{sec:mixed-energy}
\begin{figure}
    \centering
\subfig{a}{\includegraphics[width=\columnwidth]{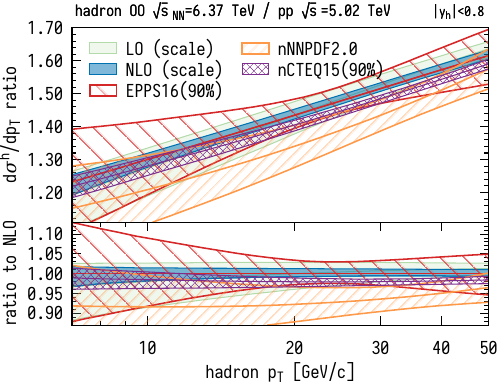}}
\subfig{b}{\includegraphics[width=\columnwidth]{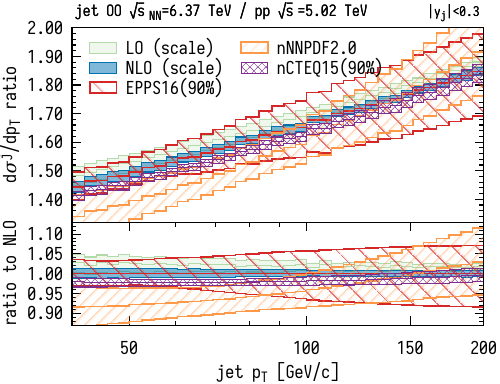}}
    \caption{No-energy-loss baseline  for the ratio of hadron (a) and jet (b) spectra in OO collisions at $\sqrt{s_{NN}}=6.37\,\text{TeV}$ with that in $pp$ at $\sqrt{s}=5.02$~TeV.
    Scale uncertainties (solid green and blue) and nPDF uncertainties (pattern) are shown.
    }
    \label{fig:pQCDjO}
\end{figure} 
For the standard nuclear modification factor, the theoretical no-energy-loss baseline differs from unity due to nuclear PDFs~\cite{Arneodo:1992wf}, so it is necessary to compare measurements to theoretical calculations to establish the presence of (small) final-state effects~\cite{,Huss:2020dwe}.
In this section we propose that energy loss in OO collisions could be discovered by measuring the ratio of 
jet or hadron spectra in OO to $pp$ spectra at a different nearby energy, and comparing this to 
the nPDF baseline in the absence of energy loss.
In \Fig{fig:pQCDjO} we show the pQCD baseline for the ratios of hadron and jet cross-sections in minimum bias OO collisions at $6.37\,\text{TeV}$ (per nucleon-nucleon pair) divided by spectra in $pp$ collisions at 
$5.02\,\text{TeV}$.

We show  nPDF uncertainties for EPPS16~\cite{Eskola:2016oht}, nNNPDF2.0~\cite{AbdulKhalek:2020yuc} and the recent nCTEQ15(WZSIH) version~\cite{Duwentaster:2021ioo}. 
EPPS16 and nCTEQ15 are consistent within the larger EPPS16 uncertainties. nNNPDF2.0 exhibits a systematic downward shift for hadrons and jets with $p_T<100\,\text{GeV}$. 
New nCTEQ15 nPDFs including inclusive hadron production and LHC $W/Z$ data have very stringent error bands (below 3\% for hadrons and 2\% for jets). It will be important to compare predictions from the upcoming iterations of EPPS21 and nNNPDF3.0 global fits with extended datasets.
Since nPDF global fits do not contain data with oxygen nuclei and hadronic  data is only available for heavy or very light nucle, interpolated nPDFs for oxygen nuclei are potentially sensitive to the assumed $A$ dependence of nPDFs. 
Complementary $p$O data would be very valuable in validating and constraining nPDF uncertainties in light nuclei \cite{Paukkunen:2018kmm}.

\section{Discussion}
In this work, we have predicted the ratio of hadron and jet spectra at two collision energies (6.37 and 5.02 TeV) in $pp$ collisions, which can be used to construct the reference spectrum for the upcoming OO run at the LHC. We used pQCD calculations and data-driven interpolation and estimated theoretical uncertainties of both methods.
The residual pQCD uncertainty on the ratio is below 2\% for hadrons and below 2\% for jets above $60\,\text{GeV}$. The uncertainty from the data-driven interpolation can be as low as 3\%, but is sensitive to the selection of data in the fits and  relies on using three energies with significant cancellation of systematic uncertainties.
It would be necessary to perform the interpolation analysis with the LHC Run 3 data at the actual collision energies once it becomes available. If both methods are considered together, the central values of pQCD and interpolated ratios deviate from each other by up to 4\% for hadrons and 2\% for jets.

Alternatively, we propose that the measured jet or hadron spectra in 6.37 TeV OO collisions can be
compared to the spectra in $pp$ collisions at a different energy by forming a mixed-energy nuclear modification ratio. In this case, there is no need to construct a $pp$ reference spectrum and the measured ratio can be compared directly to the pQCD baseline computation. 
The mixed-energy ratios have few percent scale uncertainties similar to the mixed-energy $pp$ ratios from pQCD. Nevertheless, these uncertainties are smaller or comparable to nPDF uncertainties and smaller than uncertainties we estimate for interpolating a reference from measured data.
We note that the mixed-energy nPDF baseline relies on the center-of-mass energy dependence of pQCD being correct, which can be cross-checked with data.
Different nPDFs have significant differences in their uncertainty bands, but the newest nCTEQ15 version has uncertainties reduced to a few percent.
Measurements in $p$O collisions will be especially important for providing direct constraints on the nPDFs
of oxygen~\cite{Paukkunen:2018kmm}. 
Since a $pp$ reference  measurement  is not planned at the $p$O energy in the LHC Run 3, 
reference interpolation and mixed-energy ratios may also be relevant for global nPDF fits.

The computer code to perform the MCMC fits is based on \cite{Brewer:2020och} and is publicly available at \url{https://github.com/jasminebrewer/spectra-from-MCMC}.

\section{Acknowledgements}
We thank Pit Duwentäster for nCTEQ15 nPDFs
and gratefully acknowledge valuable discussions with 
Andrea Dainese, 
Silvia Ferrario Ravasio,
Alexander Kalweit, 
Yen-Jie Lee, 
Pier Monni, 
Petja Paakkinen, 
Anne Sickles, 
Jesse Thaler, 
Marta Verweij, 
Vytautas Vislavicius, 
and Urs Wiedemann.

\bibliography{letter.bib}

\appendix

\section{MCMC fits to CMS hadron spectra\label{sec:appA}}

\begin{figure}
    \centering
    \includegraphics[width=\linewidth]{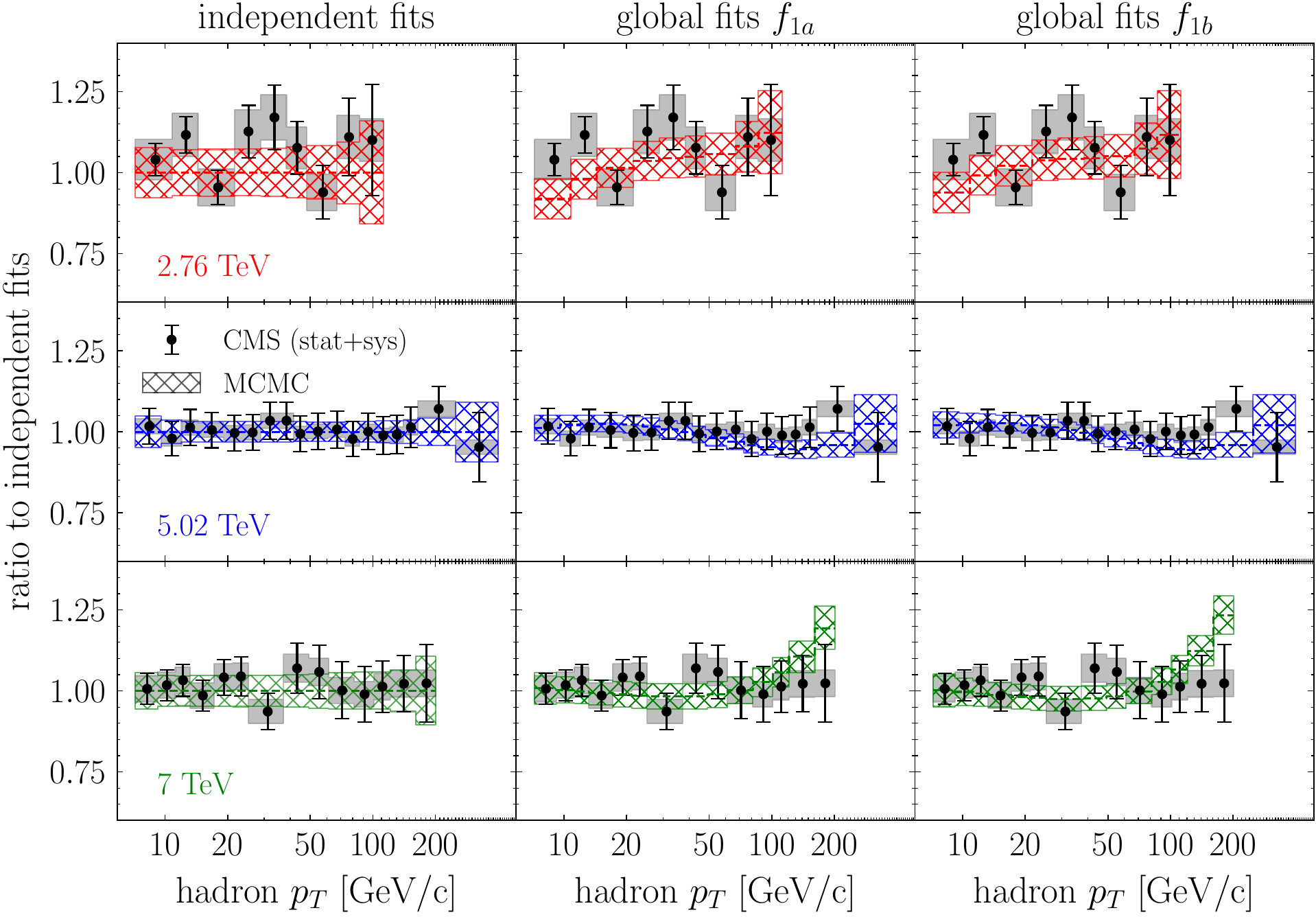}
    \caption{\label{fig:CMShadron_fits_uncorr}Fits to CMS hadron spectra at $\sqrt{s}=2.76$, 5.02, and 7~TeV~\cite{CMS:2012aa,CMS:2016xef,CMS:2011mry}, anchoring the interpolation for the results for uncorrelated systematic uncertainties shown in \Fig{fig:CMS}. The left panel shows independent fits to the spectrum at each energy.
    The center and right panels show results for global fits $f_{1a}$ and $f_{1b}$ discussed in the main text. For visual clarity, all data and fits are normalized to the median value of the independent fit at that energy. Uncorrelated (statistical + systematic) uncertainties are shown in bars and luminosity uncertainties are grey boxes. Hatched bands indicate the 68\% confidence interval of the MCMC fits.}
\end{figure}

\begin{figure}
    \centering
    \includegraphics[width=\linewidth]{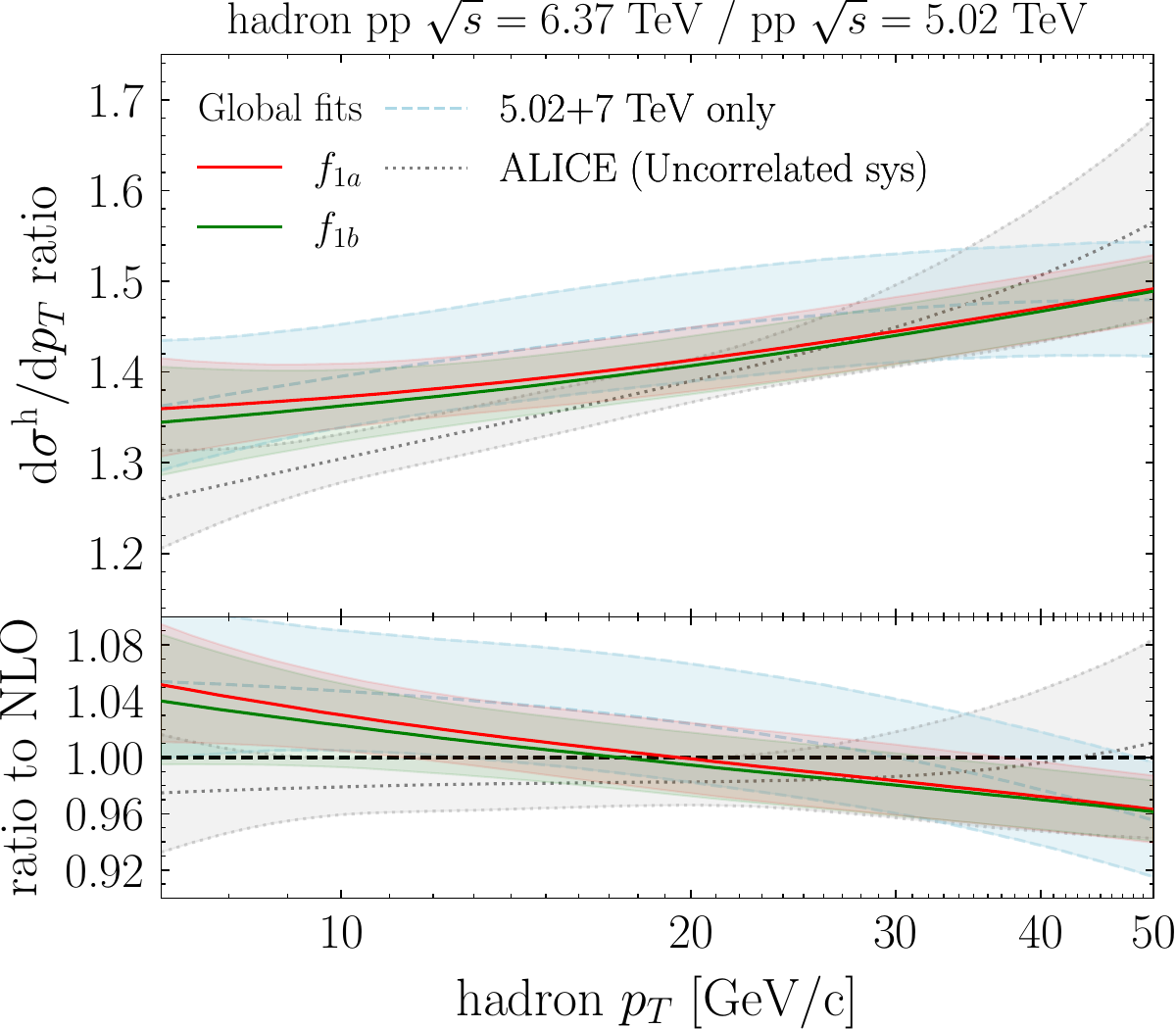}
    \caption{Ratio of inclusive hadron spectra in $pp$ collisions at 6.37 and 5.02~TeV (upper panel) and ratio to pQCD NLO (lower panel) for MCMC fits to CMS measurements. Bands show 68\% confidence intervals around the median for $f_{1a}$ (red) and $f_{1b}$ (green) fit forms. Fits using only 5.02 and 7~TeV data are shown in light blue. Fits to ALICE hadron spectra using $f_{1a}$ from \Fig{fig1}c are shown in grey. In all curves systematic uncertainties are taken to be fully uncorrelated. Note the larger scale in the lower panel relative to the main text.}
    \label{fig:CMS}
\end{figure}

Here we perform MCMC global fits to CMS hadron data at $\sqrt{s}=2.76$, 5.02 and 7~TeV. The measurements at 2.76 and 5.02~TeV are for $|\eta|<1$ \cite{CMS:2012aa,CMS:2016xef} while at 7~TeV it is for $|\eta| < 2.5$ \cite{CMS:2011mry}. We apply a correction factor of 0.9705 to the 7~TeV spectrum that was obtained from the ratio of average multiplicity in $|\eta|<1$ to $|\eta|<2.5$ in the measured pseudorapidity distribution \cite{CMS:2010tjh}. We use the inelastic hadronic cross-sections 61.8, 67.6, and 70.9~mb to convert $\sqrt{s}=2.76$, 5.02, and $7$~TeV spectra to cross-sections \cite{Loizides:2017ack}.
Figure~\ref{fig:CMShadron_fits_uncorr} shows fits to CMS hadron data with fully uncorrelated systematic uncertainties.
We note that the dispersion in the data is large compared to the luminosity and statistical uncertainties only, and therefore assuming all systematic uncertainties to cancel gives high $\chi_\nu^2 \approx 15.3$, indicating a poor fit quality. For that reason we show in \Fig{fig:CMS} only results for fully uncorrelated systematic uncertainties ($\chi^2_\nu=1.08$ and $1.13$ for $f_{1a}$ and $f_{1b}$ respectively).
We note that the interpolation using CMS data suggests a somewhat flatter $p_T$-dependence of the $6.37/5.02$~TeV ratio than comparable fits to the ALICE data from the main text (overlaid in grey), though they are consistent within large uncertainties.

\end{document}